\documentclass[12pt,preprint]{aastex}

\bibpunct{(}{)}{,}{a}{}{,}

\shorttitle{Ly$\alpha$ blobs at $z$=3.1}
\shortauthors{Matsuda et al.}

\begin{document}


\title{A Keck/DEIMOS Spectroscopy of Ly$\alpha$ Blobs at Redshift $z=3.1$\altaffilmark{1,2}}


\author{Yuichi Matsuda      \altaffilmark{3},
        Toru Yamada         \altaffilmark{4},
        Tomoki Hayashino    \altaffilmark{5},
        Ryosuke Yamauchi    \altaffilmark{5},
        Yuki Nakamura       \altaffilmark{5}
}


\email{matsdayi@kusastro.kyoto-u.ac.jp}

\altaffiltext{1}{The data presented herein were obtained at the W.M. Keck Observatory, which is operated as a scientific partnership among the California Institute of Technology, the University of California and the National Aeronautics and Space Administration. The Observatory was made possible by the generous financial support of the W.M. Keck Foundation.}

\altaffiltext{2}{Based on data collected at Subaru Telescope and in part obtained from data archive at Astronomical Data Analysis Center, which are operated by the National Astronomical Observatory of Japan.}

\altaffiltext{3}{Department of Astronomy, Kyoto University, Sakyo-ku, Kyoto 606-8502, Japan; matsdayi@kusastro.kyoto-u.ac.jp}

\altaffiltext{4}{National Astronomical Observatory of Japan, Mitaka, Tokyo 181-8588, Japan; yamada@optik.mtk.nao.ac.jp}

\altaffiltext{5}{Research Center for Neutrino Science, Graduate School of Science, Tohoku University, Aramaki, Aoba, Sendai 980-8578, Japan;  haya@awa.tohoku.ac.jp; yamauchi@awa.tohoku.ac.jp; y.naka@awa.tohoku.ac.jp}


\begin{abstract}

 We present the results of an intermediate resolution ($\sim 2$ \AA ) 
spectroscopy of a sample of 37 candidate Ly$\alpha$ blobs and emitters 
at redshift $z=3.1$ using the DEIMOS spectrograph on the 10 m Keck 
telescope. The emission lines are detected for all the 37 objects and 
have variety in their line profiles. The Ly$\alpha$ velocity widths (FWHM) 
of the 28 objects with higher quality spectra, measured by fitting a 
single Gaussian profile, are in the range of $150 - 1700$ km s$^{-1}$ and 
correlate with the Ly$\alpha$ spatial extents. All the 12 Ly$\alpha$ blobs 
($\ge 16$ arcsec$^2$) have large velocity widths of $\ga 500$ km 
s$^{-1}$. While there are several possible physical interpretations of the 
Ly$\alpha$ velocity widths (motion of gravitationally-bound gas clouds, 
inflows, merging of clumps, or outflows from superwinds), the large velocity 
widths of the Ly$\alpha$ blobs suggest that they are the sites of massive 
galaxy formation. If we assume gravitationally-bound gas clouds, the 
dynamical masses of the Ly$\alpha$ blobs are estimated to be 
$\sim 10^{12} - 10^{13}$ M$_{\odot}$. Even for the case of outflows, 
the outflow velocities are likely to be the same order of the rotation 
velocities as inferred from the observational evidence for local starburst 
galaxies. 

\end{abstract}





\section{INTRODUCTION}

 Ly$\alpha$ blobs (LABs) are diffuse, large and radio-quiet Ly$\alpha$
nebulae often discovered in galaxy overdense regions at high redshifts
(Keel et al. 1999; Steidel et al. 2000, hereafter S00; Francis et
al. 2001; Palunas et al. 2004; Matsuda et al. 2004, hereafter
M04). Recent observations revealed that a significant fraction of LABs
has large luminosity of infrared dust continuum emission (Chapman et
al. 2001; Smail et al. 2003; Dey et al. 2005; Geach et al. 2005; Colbert
et al. 2006) and that some LABs have large ($\ga 1000$ km s$^{-1}$)
Ly$\alpha$ velocity widths (S00; Ohyama et al. 2003; Bower et al. 2004;
Dey et al. 2005; Wilman et al. 2005). Although their physical origins
are still unclear, three models have been proposed, (i) photo-ionization
by hidden AGNs or starbursts (S00; Chapman et al. 2001; Busu-Zych \&
Scharf 2004; Furlanetto et al. 2005), (ii) Ly$\alpha$ cooling radiation
by gravitational heating (Haiman et al. 2000; Fardal et al. 2001; 
Dijkstra et al. 2005), and (iii) shock heating by starburst driven 
galactic superwinds (Taniguchi \& Shioya 2000; Ohyama et al. 2003; Mori 
et al. 2004).

 On the other hand, a large number of compact Ly$\alpha$ emitters 
(LAEs) have been discovered (Cowie \& Hu 1998; Fynbo et al. 2003; 
Ouchi et al. 2003; Shimasaku et al. 2004; Hayashino et al. 2004). 
Their compact morphology suggests that they are small building 
blocks of galaxies (Pascarelle et al. 1998). Their faint rest frame 
optical luminosity suggests that they are young or have small 
stellar masses (Yamada et al. 2001). They do not show large infrared 
luminosity (De Breuck et al. 2004). They show narrow (a few hundred 
km s$^{-1}$) Ly$\alpha$ velocity widths (Hu et al. 2004; Dawson et 
al. 2004; Venemans et al. 2005). Giant LABs such as discovered in 
S00 are known to be $20 - 40$ times larger than typical compact 
LAEs. However, M04 revealed the existence of a large number of 
slightly smaller and fainter blobs that show rather continuous 
distribution in Ly$\alpha$ spatial extents between giant LABs and 
compact LAEs. It is thus also important to understand the relation 
between LABs and compact LAEs in order to reveal the nature of LABs.

 In this letter, we report the results of an intermediate resolution 
($\sim 2$ \AA ) spectroscopy of Ly$\alpha$ sources ranging from giant 
LABs to compact LAEs at $z=3.1$ in the SSA22 field. Matsuda et al. (2005) 
previously obtained 56 spectra for LAEs in the same field, but the depths 
and the spectral resolution were insufficient to examine their line 
profiles. We assume an $\Omega_{\rm M} = 0.3$, $\Omega_{\Lambda} =0.7$ 
universe with $H_0 = 70$ km s$^{-1}$ Mpc$^{-1}$ ($1''.0$ corresponds 
to 7.6 kpc of physical length at $z=3.1$).

\section{OBSERVATIONS}

The spectroscopic targets were chosen from candidate Ly$\alpha$ sources
identified in M04. We briefly describe the selection criteria below. We used 
narrow-band ($NB497$; 4977\AA /77\AA) and broad-band images in the SSA22 field 
($\alpha = 22^{\rm h}17^{\rm m}34^{\rm s}.0$, $\delta = +00^{\rm o}17'01''$, 
[J2000]) obtained with the Suprime-Cam (Miyazaki et al. 2002) on the 8.2 m 
Subaru telescope. The average stellar profile of the images has a full 
width at half maximum (FWHM) of $1''.0$. The object 
detection was made on a continuum subtracted narrow-band ($NB_{\rm corr}$) 
image smoothed with a Gaussian kernel with a FWHM of $1''$ using a threshold 
of emission line surface brightness of $2.2 \times 10^{-18}$ ergs s$^{-1}$ 
cm$^{-2}$ arcsec$^{-2}$. The magnitudes and colors are calculated by the 
{\it same} isophotal apertures determined on the smoothed $NB_{\rm corr}$ 
image with the same isophotal limit as the detection threshold. We selected 
277 candidate Ly$\alpha$ sources, which satisfy the following criteria: 
(1) $NB497<25$ AB mag, and (2) observed equivalent width $EW_{\rm obs}>80$ 
\AA\ (or Ly$\alpha$ luminosity L$({\rm Ly}\alpha)> 1.7 \times 10^{42}$ erg 
s$^{-1}$ at $z=3.1$) (Fig. 1). We denote them as No. 001 through No. 277 in 
order of isophotal areas. The largest 35 candidates with isophotal areas 
$\ge 16$ arcsec$^2$ (or $\ge 900$ kpc$^2$ at $z=3.1$) are identical with 
the LABs presented in M04. The candidates lying near the solid curve in 
Figure 1 (the expected values for point sources) are not resolved in the 
narrow-band image. We will call the smallest 129 candidates (No. 149 -- 
No. 277) with the isophotal areas $< 8$ arcsec$^2$, compact LAEs.

 We carried out spectroscopic observations of 37 out of the 277 candidates 
using the DEIMOS spectrograph (Faber et al. 2003) on the 10 m Keck II 
telescope in August 14--18 2004 (UT). The 37 targets are widely distributed 
in isophotal areas (see Fig. 1). The field of view of a slit mask is about 
$5'\times 16'$. We used the 900 l/mm grism and the GG455 filter. The typical 
wavelength coverage is $\sim 4550 - 7500$ \AA . The slit widths for all 
objects are $1''.0$. The slit lengths range from $6''$ to $32''$ in order 
to obtain sky outside of their Ly$\alpha$ spatial extents. The spectral 
resolution is 2.2 \AA\ at 5000 \AA\ or a velocity resolution of 125 km 
s$^{-1}$. The pixel sampling is 0.44 \AA\ in wavelength and $0''.12$ in 
spatial direction. We used one slit mask in this observing run. Total 
exposure time was 7 hours with each 30 minutes exposure. The seeing ranged 
from $0''.8$ to $1''.3$. We reduced the data using spec2d pipeline. For 
analysis of integrated line profiles discussed in this letter, we made one 
dimensional spectra combining $1''.1 \times 1''.0$ around emission line 
peaks. While extended emission lines were clearly detected for some objects, 
a study of the faint outskirts with lower signal-to-noise ratio (S/N) needs 
further careful analysis and the results will be presented elsewhere in the 
future. The spectra are not flux calibrated.

\section{RESULTS}

 We detected emission-lines for all the observed 37 spectra with peaks above 
S/N of 3 per spectral resolution. The emission lines have variety in their 
line profiles (Fig. 2). For example, No. 004 has double peaks. The both 
blueshifted and redshifted emission lines have clear asymmetry. When we use 
statistics of line asymmetry\footnote{The ``wavelength ratio'' is defined as 
$a_{\lambda} = (\lambda_{10,r}-\lambda_{p})/(\lambda_{p}-\lambda_{10,b})$ 
and the ``flux ratio'' as $a_f = \int_{\lambda_{p}}^{\lambda_{10,r}} d\lambda 
/ \int_{\lambda_{10,b}}^{\lambda_{p}} d\lambda $, where $\lambda_{p}$ is the 
wavelength of the peak of the emission line, and $\lambda_{10,b}$ and 
$\lambda_{10,r}$ are the points at which the line flux exceeds 10\% of the 
peak on the blue side and the red side respectively.}, $a_{\lambda}$ and 
$a_f$, proposed by Rhoads et al. (2003), the lines have values of 
$a_{\lambda}=0.33$, $a_f=0.48$ (the blueshifted line), and $a_{\lambda}=3.6$, 
$a_f=4.5$ (the redshifted line). These values show that the blueshifted line 
has a steeper drop on the red side than the blue side, and the redshifted 
line has opposite asymmetry. Thus, it is reasonable to interpret them as a 
relatively broad emission line with a single narrow absorption. No. 031 has 
line profile similar to that of No. 004. Similar line profiles were also 
reported for LAB2 in the same field (Wilman et al. 2005), $z\sim 3.1$ LAEs 
around a radio galaxy (Venemans et al. 2005) and radio galaxies at 
$z\sim 2.5-4$ (Wilman et al. 2004). No. 030 has a single peaked emission line 
with a strong absorption on the blue side ($a_{\lambda}=2.2$, $a_f=1.9$). 
No. 103, 151 and 241 have symmetric single narrow emission lines 
($a_{\lambda}=0.7-1.3$, $a_f=0.7-1.0$). Although these lines do not show 
clear asymmetry often seen in Ly$\alpha$ emission lines at higher redshifts
(e.g. Hu et al. 2004; Dawson et al. 2004), this may be due to the fact that 
the effect of absorptions by intergalactic \ion{H}{1} clouds on the blue 
side of Ly$\alpha$ line is weaker at $z\sim 3$ (Songaila et al. 2004). These 
spectra show no other emission lines to suggest that the identified lines are 
not Ly$\alpha$. The spectral resolution of 2.2 \AA\ is sufficient to resolve 
foreground [\ion{O}{2}]$\lambda \lambda 3726,3729$ doublets if there are. 
Only one object (No. 237) turns out to be a $z=0.33$ [\ion{O}{2}] emitter, 
as it has very narrow double peaked emission lines with 
the separation of $\sim 4$ \AA , and [\ion{O}{3}]$\lambda \lambda 4959,5007$ 
and H$\beta$ at longer wavelength (Fig. 3). No. 053 and 070 have the 
broadest emission lines and show \ion{C}{4} emission lines at longer 
wavelength. Their broad Ly$\alpha$ emission lines are likely to be due to 
AGNs in the objects. 

 We examined the {\it entire} velocity widths of the Ly$\alpha$ emission 
lines by fitting a single Gaussian function. We used only emission line part 
(i.e. within $\sim \pm 4$ times the dispersion of the Gaussian function), and 
masked possible absorption lines as shown in Figure 2. If the spectra show 
double peaks, we masked the region between the peaks. We then studied the 
distribution of Ly$\alpha$ velocity widths of 28 objects with peaks above 
S/N of 6 per spectral resolution without 2 AGNs. They have the velocity 
widths (FWHM) in the range of $150 - 1700$ km s$^{-1}$, with 
a median of 550 km s$^{-1}$. The widths are corrected for the instrumental 
resolution. The 12 LABs have the widths of $\ga 500$ km s$^{-1}$, with 
a median of 780 km s$^{-1}$. The 7 compact LAEs have the widths of 
$\la 700$ km s$^{-1}$, with a median of 280 km s$^{-1}$. Venemans et al. 
(2005) reported that $z\sim 3.1$ LAEs around a radio galaxy have the widths 
in the range of $120 - 800$ km s$^{-1}$, with a median of 260 km s$^{-1}$. 
These values are very similar to those of our compact LAEs. We plot the 
Ly$\alpha$ velocity widths and the Ly$\alpha$ isophotal areas in Figure 4. 
There is a correlation between the velocity widths and the isophotal areas. 
A Spearman rank correlation analysis shows that the correlation has 
$r_{s}=0.59$ and significance level of 99.9\%. The uncertainties of velocity 
widths by the fit are shown in Figure 4. We also estimated the uncertainties 
of the isophotal areas as follows. We put $40''\times40''$ $NB_{\rm corr}$ 
images of each object at 100 random positions on the original $NB_{\rm corr}$ 
image and re-measured the isophotal areas. Note that the derived 
uncertainties of the isophotal areas may be slightly overestimated because 
the re-measured isophotal areas were affected by combinations of noise from 
the thumbnail images of the objects and the original $NB_{\rm corr}$ image.

\section{DISCUSSION}

 We found the correlation between the Ly$\alpha$ velocity widths and the 
Ly$\alpha$ isophotal areas and that all the LABs have the widths of 
$\ga 500$ km s$^{-1}$. There are several possible physical 
interpretations of the Ly$\alpha$ velocity widths; motion of 
gravitationally-bound gas clouds, inflows, merging of clumps, or outflows 
from superwinds. 

 We consider the first case that the Ly$\alpha$ velocity widths are due to 
motion of gravitationally-bound gas clouds within collapsed dark matter halos. 
If we assume random motions in a singular isothermal sphere 
($\rho \propto r^{-2}$), the dynamical mass can be estimated to be 
$M_{\rm dyn} = 3 \sigma^2 \, r/G$, or 
\begin{equation}
 M_{\rm dyn} = 5.2 \times 10^{12} M_{\odot}~
        \left(\frac{\sigma}{500\,{\rm km~s^{-1}}}\right)^2~
        \frac{r}{30\,{\rm kpc}}
\end{equation}
where $\sigma$ is one dimensional velocity dispersion, $G$ is the 
gravitational constant and $r$ is the radius of Ly$\alpha$ nebulae. When 
we use the radius of a circle with area which is equivalent to the 
Ly$\alpha$ isophotal area, the dynamical masses of the 12 LABs are 
estimated to be $5\times 10^{11} - 2 \times 10^{13}$ M$_{\odot}$ (Fig. 4a). 
The mass estimation suggests that LABs already have their massive halos 
comparable to those of present day massive galaxies. If we consider inflows 
or merging of clumps, the gas clouds or clumps may have not yet been 
accelerated enough in the gravitational potential of dark matter halos. 
In these cases, the mass estimation may give lower limits.

 We consider the next case that the Ly$\alpha$ velocity widths are due to 
gas outflows from superwinds. In this case, the velocity widths may not 
represent the masses of the objects. Heckman et al. (2000) reported that 
there is no tight correlation between the rotation velocities and the 
outflow velocities for low redshift starburst galaxies. Martin (2005), 
however, recently revisited the issue and revealed a correlation with a 
larger sample. If the same correlation holds at high redshifts, Ly$\alpha$ 
sources with larger velocity widths would have larger dynamical masses. 
If we assume outflows from a central starburst, we can also estimate the 
characteristic time-scales of star-formation by their spatial extents and 
typical outflow speeds. If we divide $2 r$ by the Ly$\alpha$ velocity 
widths, the time-scales are estimated to be in the range of 
$3\times 10^7 - 10^8$ yr (Fig. 4b). Note that the time-scales are only upper 
limits for unresolved compact LAEs. Geach et al. (2005) suggested that the 
LABs have average star-formation rate of $\sim 10^3$ M$_{\odot}$ yr$^{-1}$ by 
their sub-mm observations. If the LABs have continuous star formation at this 
rate for several $\times 10^7$ yr, they would have large stellar masses of 
several $\times 10^{10}$ M$_{\odot}$. 

 Thus the large velocity widths of LABs imply that they are indeed hosted 
in the massive systems. Their large spatial extents, large Ly$\alpha$ 
luminosity, and rather chaotic structures seen in their line profiles, as 
well as high sub-mm detection rate, also support that they are in active 
forming phase.

\acknowledgments

 We thank the anonymous referee for useful comments which have significantly 
improved the paper. We thank the staff of the Subaru Telescope and the Keck 
Telescope for their assistance with our observations. We thank K. Ohta, 
T. Totani, and M. Mori for useful discussions. This work is supported by the 
Grant-in-Aid for the 21st Century COE "Center for Diversity and Universality 
in Physics" from the Ministry of Education, Culture, Sports, Science and 
Technology (MEXT) of Japan. The research of T.Y. is partially supported by the 
Grant-in-Aid for scientific research from the MEXT (14540234 and 17540224). 
The analysis pipeline used to reduce the DEIMOS data was developed at 
UC Berkeley with support from NSF grant AST-0071048.






\clearpage

\begin{figure}
\figurenum{1}
\epsscale{1.0}
\plotone{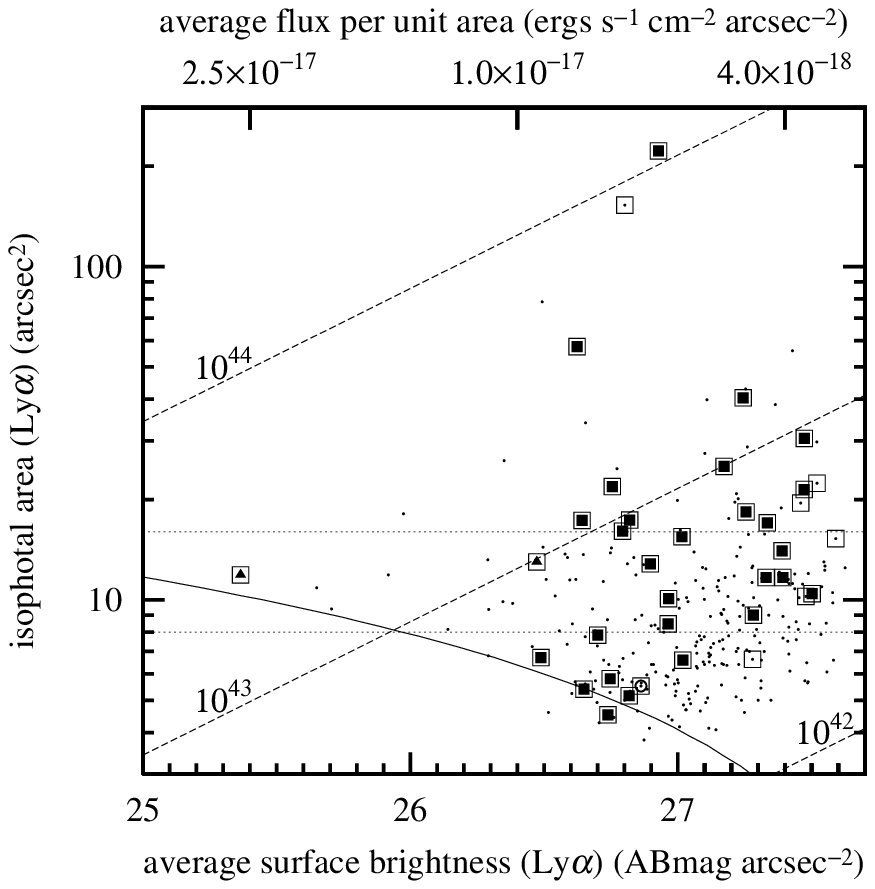}
\caption{ Distribution of Ly$\alpha$ isophotal area and Ly$\alpha$ 
average surface brightness for candidate Ly$\alpha$ sources. 
The dots show the 277 candidates. The large open squares show the 
37 spectroscopic targets. The filled squares show the 28 Ly$\alpha$ sources 
which show Ly$\alpha$ emission lines with S/N$>$6 per spectral resolution 
of 2.2 \AA . The filled triangles show AGNs at $z=3.1$. The open circle 
shows a [\ion{O}{2}] emitter at $z=0.33$. The long dashed lines correspond to 
the Ly$\alpha$ luminosity of 10$^{44}$, 10$^{43}$ and 10$^{42}$ ergs 
s$^{-1}$. The short dashed horizontal lines show the sample thresholds of 
the isophotal areas of 16 arcsec$^2$ for LABs and 8 arcsec$^2$ for compact 
LAEs.}
\end{figure}

\clearpage

\begin{figure}
\figurenum{2}
\epsscale{0.75}
\plotone{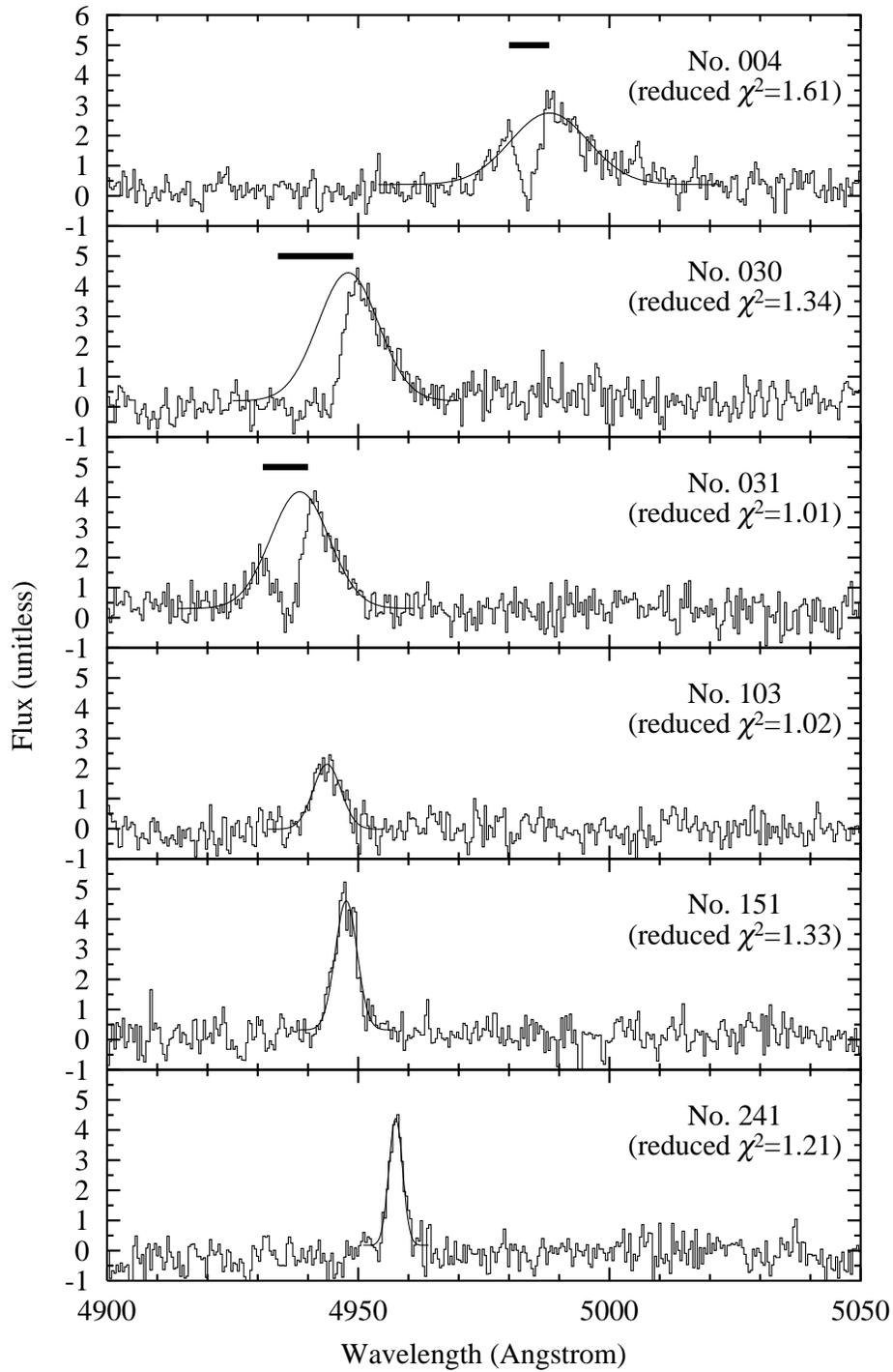}
\caption{ Examples of 1 dimensional spectra of Ly$\alpha$ sources. The solid 
curves show the best fit Gaussian profiles. The thick bars show masked 
regions.}
\end{figure}

\clearpage

\begin{figure}
\figurenum{3}
\epsscale{0.75}
\plotone{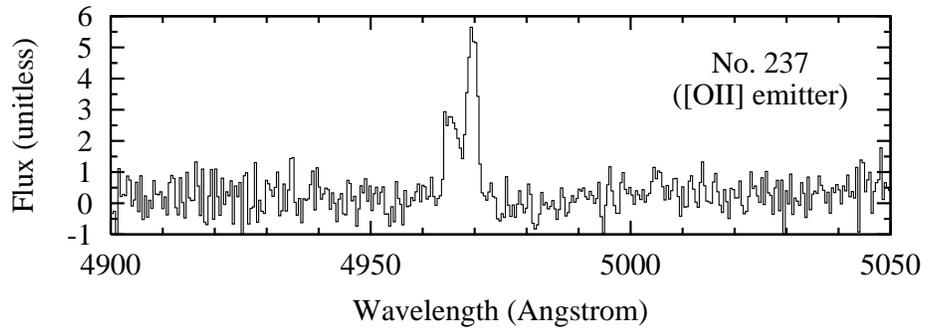}
\caption{ A spectrum of a foreground [\ion{O}{2}] emitter at $z=0.33$ in 
the spectroscopic sample. The spectral resolution of 2.2 \AA\ is sufficient 
to resolve [\ion{O}{2}]$\lambda \lambda 3726,3729$ doublets.}
\end{figure}
\clearpage

\begin{figure}
\figurenum{4}
\epsscale{1.0}
\plotone{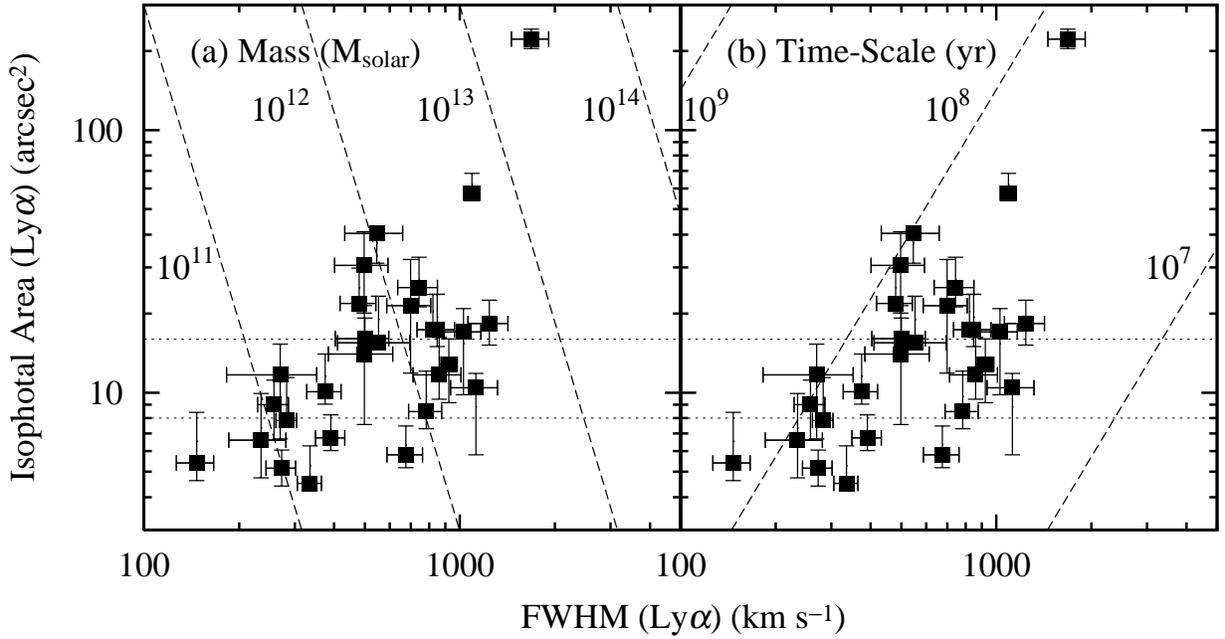}
\caption{ Distribution of Ly$\alpha$ isophotal area and Ly$\alpha$ velocity 
width (FWHM). The filled squares show 28 Ly$\alpha$ sources with the 
Ly$\alpha$ emission line peaks above S/N of 6 per spectral resolution of 
2.2 \AA . The short dashed horizontal lines show the sample thresholds of 
the isophotal areas of 16 arcsec$^2$ for LABs and 8 arcsec$^2$ for compact 
LAEs. ({\it left panel} ) The long dashed lines show the estimated dynamical 
masses of $10^{11}$, $10^{12}$, $10^{13}$, $10^{14}$ M$_{\odot}$ assuming 
motion of gravitationally-bound gas clouds. ({\it right panel} ) The dashed 
lines show the estimated time-scales of star-formation of $10^7$, $10^8$, 
$10^9$ yr assuming outflows from superwinds.}
\end{figure}

\end{document}